# Novel brain biomarkers of obesity based on statistical measurements of white matter tracts

José Gerardo Suárez-García[a], María Isabel Antonio-de la Rosa[a], Nora Coral Soriano-Becerril[b], Javier M. Hernández López[a], Benito de Celis-Alonso[a*]

[a] Faculty of Physical and Mathematical Sciences, Benemérita Universidad Autónoma de Puebla (BUAP), Puebla, Mexico
[b] Instituto Mexicano del Seguro Social (IMSS), Puebla, Mexico

*Correspondence author
E-mail address: bdca@fcfm.buap.mx (B. de Celis-Alonso)

# Abstract

**Objective:** Novel brain biomarkers of obesity were sought by studying statistical measurements on fractional anisotropy (FA) images of different white matter (WM) tracts from subjects with specific demographic characteristics.

**Methods:** Tract measurements were chosen that showed differences between two groups (normal weigh and overweight/obese) and that were correlated with their BMI. From these measurements, a simple and novel process was applied to select those that would allow the creation of models to quantify and classify the state of obesity of individuals. The biomarkers were created from the tract measurements used in the models.

**Results:** Some positive correlations were found between WM integrity and BMI, mainly in tracts involved in motor functions. From this result, neuroplasticity in motor tracts associated with obesity was hypothesized. Two models were built to quantify and classify obesity status, whose regression coefficients formed the novel proposed obesity-associated brain biomarkers.

**Conclusion:** A process for the selection of tract measurements was proposed, such that models were built to determine the obesity status of subjects individually. From these models, novel brain biomarkers associated with obesity were created. These allow the generation of new knowledge and are intended to be a future tool in the clinical environment for the prevention and treatment of brain changes associated with obesity.

**Significance:** After studying subjects with specific demographic characteristics, results opposed those usually reported in literature. These consisted of positive correlations between WM integrity and obesity mainly in tracts involved in motor functions, suggesting neuroplasticity in these tracts. Novel brain biomarkers of obesity were also proposed, formed by the regression coefficients involved in precise models of quantification and classification of obesity status. All this allows the generation of new knowledge and its probable subsequent clinical application.

# Highlights

• Positive correlations between WM integrity and obesity particularly in motor tracts were found.

• A simple and novel process for choosing tract measurements for the construction of regression models was proposed.

• Two models were created for the quantification and classification of obesity status of individual subjects.

• Novel brain biomarkers associated with obesity were constructed from the tract measurements involved in the created models.

• Findings allowed to hypothesize possible neuroplasticity in motor tracts associated with obesity.

# 1. Introduction

Obesity is a major health problem worldwide. In 2022, 2.5 billion adults in the world (which corresponds to 43% of the total) were overweight, including over 890 million adults who were living with obesity [1]. It is known that obesity induces a myriad of pathological alterations that include between others: neuroinflammation, vascular damage, metabolic imbalances, is a precursor of some types of cancers and blood–brain barrier disruption [2].

In the neurological context, several works have found associations between obesity and the structure and function of the gray matter (GW) and white matter (WM) of the brain analyzing multi-modal magnetic resonance imaging (MRI), positron emission tomography (PET), and single-photon emission computed tomography (SPECT) [3]. Regarding WM, diffusion tensor imaging (DTI) is one MRI modality that allows to study integrity and coherence of WM tracts through a scalar measurement called fractional anisotropy (FA). It measures the water diffusivity along the axons that form the WM tracts with high FA values representing highly organized and normally myelinated axon structures [4]. On the other hand, reduced FA values can be interpreted as a loss of coherence in the main preferred diffusion direction, resulting in a deficit in white matter microstructure and integrity. FA has become the most used DTI measurement to quantify WM properties, specifically in voxel-based analysis (VBA) and tract-based and spatial statistic (TBSS) [5].

Most of the works studying FA have found that adults with a higher BMI have poorer WM integrity, indicating that the fiber tracts connecting various brain regions are less efficient and slower during the neural transmission and information processing [6]. In a recent work, Dietze et al. [7] included 51 studies in a voxel based meta-analysis. They demonstrated an association between reduced FA and obesity in the genu and splenium of the corpus callosum, middle cerebellar peduncles, anterior thalamic radiation, corticospinal projections, and cerebellum. In other work, Chen et al. [8] conduct a coordinate-based meta-analysis on 5 studies. Their findings showed that overweight or obese individuals exhibited reduced FA compared to subjects with normal weight, in the right superior longitudinal fasciculus, the splenium of the corpus callosum, the right median network, and cingulum. Besides, Okudzhava et al. [9] conducted a systematic review considering 31 studies, finding that the majority reported decreased FA associated with elevated BMI.

Despite all the proof presented of decreased FA with increased BMI, other studies found no significant differences between FA values and obesity or even positive associations between both in some tracts [5]. Three works of this kind are worth mentioning. Verstynen et al. [10] found that the FA of WM voxels parametrically decreased with increasing BMI. However, several clusters with strong positive relationships were also reported in the middle of the cerebellar peduncle and the fiber pathways near the colliculus. Birdsill et al. [11] found that higher waist circumference was associated with higher FA in the posterior white matter bilaterally. Based on their results, the authors speculated that possible degenerative processes could account for the positive association found, such as myelin repair, loss of long white matter tracts with sparing of short interneurons, and a lack of reorganization. Finally, Rahmania et al. [12] investigated whether gender differentially affects the relationship between BMI and WM structural connectivity. In contrast to men, women showed a relationship between higher BMI and increased connectivity in several WM tracts, including the bilateral frontoparietal cingulum, bilateral reticulospinal, dentatorubrothalamic, corticopontine tracts, the left corticospinal tract and the tapetum area of corpus callosum.

Different works have reported findings at the group level and not individually. This implies that those results provide statistical information about the groups studied, but this information does not always allow obtaining precise information for individual subjects. Therefore, biomarkers cannot be proposed based on this information, since their purpose is to associate specific characteristics with individual subjects. Nevertheless, this characterization has been attempted by creating models using linear regression, neural networks, etc., that combine different quantitative characteristics of the brain (structural or functional) to predict the obesity status of individual subjects [2,3,13,14]. The objective of these models was not to use a complex methodology to diagnose the obesity status of a subject. This can be easily described for example by BMI, which is easily measured with a scale and a meter. Their objective was to discover relationships between brain parameters measured, and the already known obesity status of an individual. Subsequently and based on those relationships, authors proposed obesity biomarkers for the general population.

In the present work, obesity biomarkers were sought by studying a set of subjects from a specific demographic group. The biomarkers were formed by a combination of statistical measurements made on DTI tracts and were based on the FA values. From here on these values will be named by us as tract measurements. To carry out this project, the protocol was divided into two parts. In the first, different tract measurements were chosen when they met two conditions: Presenting significant differences between the two study groups (NW and OB) (condition one) and being significantly correlated with the subjects BMI (condition two). This could be considered as the extraction of information at group level (since it was obtained from comparing two study groups). In the second part with the chosen tract measurements, two regression models were built to predict the obesity status of the subjects. One quantifying the BMI and the other classifying subjects into the obese (OB) or normo-wight (NW) groups. This step could be considered as obtaining information at individual level (since the obesity status was predicted individually for each subject). Then, obesity biomarkers were constructed by combining the tract measurements used in the models. This work had two hypotheses. The first was that, compared to that reported usually by other works where they study a wide range of ages and mixed genders, in this work different findings at the group level will be obtained studying a set of subjects with specific demographic variables. And the second, using the previous findings, new biomarkers associated with obesity will be constructed from tract measurements involved in the predicting models. Both aims look to provide new knowledge on alterations in WM associated with obesity.

## 2. Methodology

In the first part of the work, tract measurements were obtained by calculating different statistics on the FA data of each WM tract. Then, among all the measurements, those that met two conditions were chosen. In the second part, considering the previously chosen tract measurements, two regression models were built. A novel methodology was proposed to select tract measurements that would allow the creation of accurate models. Finally, new obesity biomarkers were proposed. Unless otherwise stated, algorithms were developed in MATLAB R2023b, using a conventional computing system (Intel Core i7-12700H, NVIDIA GeForce RTX 3070 Ti, 32 GB RAM).

## 2.1 Database

The open-access database Amsterdam Open MRI Collection (AOMIC) was used for this study, which consisted of three large-scale datasets with high-quality, multimodal 3T MRI data and detailed demographic and psychometric data from a large set of healthy participants. Of the three datasets, the so-called "ID1000" was studied [15], which is a large representative dataset of the general population. AOMIC contains both raw data as well as preprocessed data from well-established preprocessing and quality control pipelines. Among the different available MRI modalities, preprocessed diffusion-weighted MRI (DWI) were considered in the present work, from which data derived from the original raw data consisting of fractional anisotropy (FA) maps were studied. Basic information about the database will be summarized below, although a more detailed description can be found on the database website [16], and in a paper published in Nature Scientific Data [17].

### 2.1.1 MRI Scanning protocol

Based on the description included in the database page, data from ID1000 dataset were scanned on a Philips 3T scanner (Philips, Best, the Netherlands). On the "Intera" version using a 32-channel head coil. At the start of each scan session, a low-resolution survey scan was made, which was used to determine the location of the field-of-view. Three $T_1$-weighted scans, three diffusion-weighted scans, and one functional (BOLD) MRI scan were recorded (in that order). For all diffusion scans, the slice stack was not angled. Three scans were obtained with the SE-DWI technique with a b0 image, 32 diffusion-weighted directions, a half sphere sampling scheme, and DWI b-value equal to 1000 s/mm$^2$. Voxel size was equal to 2×2×2 mm, FOV of 224×224×120, matrix size of 112×112, 60 slices with no slice gap, TR = 6370 ms and TE = 75 ms, water-fat shift of 12,861 pixels, bandwidth equal to 33.8 Hz/pixel, flip angle of 90 degrees and with a duration of 4 minutes and 49 seconds.

### 2.1.2 DWI standardization, preprocessing and FA image computing

The following set of processes were already implemented on the database by creators. According to its description, data were converted to BIDS, including file renaming, conversion to compressed nifti, and defacing and extraction of metadata. The three DWI scans per participant, the diffusion gradient table, and b-value information were concatenated. Following this, preprocessing was applied to the data using tools from MRtrix3 and FSL. This consisted of denoising the diffusion-weighted data, removing Gibbs ringing artifacts, and performing eddy current and motion corrections. Within the eddy, a quadratic first-level and linear second-level model and outlier replacement with default parameters were used. Bias correction, brain mask extraction, and possible issues correction on the diffusion gradient table were also performed. A diffusion tensor model on the preprocessed diffusion-weighted data using weighted linear least squares with 2 iterations was fit. From the estimated tensor image, a fractional anisotropy (FA) image was computed and a map with the first eigenvectors was extracted.

### 2.1.3 Affine aligned into MNI152 standard space

In the present work, an additional affine alignment of the FA images into MNI152 standard space was performed. To this end, two FSL scripts available online (developed originally to perform TBSS) were applied [18]. The first script was tbss_2_reg, used to align all FA images to a 1x1x1mm standard space by performing nonlinear registration and considering the adult-derived target image FMRIB58_FA. The second script was tbss_3_postreg, which made nonlinear transformations to bring the images into MNI152 standard space.

### 2.1.4 Subjects included in the study

Data from 992 subjects (men and women) were available within the ID1000 dataset. According to the authors, subject sample was representative of the general Dutch population in terms of educational level (as defined by the Dutch government), with age range from 19 to 26 years, to minimize the effect of aging on any brain-related covariates. Subjects' body-mass-index (BMI) was calculated and rounded to the nearest integer. Educational level was reported on a three-point scale: low, medium, and high, based on the completed or current level of education. In the present work, to reduce the variability between the demographic characteristics of the subjects, only right-handed females with a medium level of education were considered (including upper secondary education (HAVO/VWO), basic vocational training (MBO-2), vocational training (MBO-3), and middle management and specialist education (MBO-4), based on the Dutch education system [19]) and with an BMI over 19 kg/m$^2$ (excluding subjects with low weight). In total, 160 subjects met the criteria for inclusion (**Table S1** and **S2**). The set formed by all the subjects was called $D_{all}$. From this, two subsets were created, one called $D_{norm}$, consisting of 80 subjects with 19 kg/m$^2$ ≤ IBM < 25 kg/m$^2$ (NW), and another called $D_{over}$, consisting of 80 subjects with 25 kg/m$^2$ ≤ IBM ≤ 47 kg/m$^2$ (OB, with 47 kg/m$^2$ being the highest IBM available). Information on $D_{norm}$ and $D_{over}$ subsets is shown in Table 1.

| | $D_{norm}$ | $D_{over}$ | | |
|---|---|---|---|---|
| **Subjects** | Normal<br>19 kg/m$^2$ ≤ BMI < 25 kg/m$^2$ | Overweight<br>(25 ≤ BMI < 30) | Obese<br>(30 ≤ BMI < 35) | Extremely obese<br>(35 ≤ BMI ≤ 47) |
| | 80 (total) | 49 | 19 | 12 |
| | | 80 (total) | | |
| **Age (years)** | 22.65 ± 1.66 | 22.63 ± 1.71 | | |
| **BMI (kg/m$^2$)** | 21.63 ± 1.72 | 29.55 ± 4.63 | | |

**Table 1. Information about $D_{norm}$ and $D_{over}$ subsets**. The number of subjects according to their weight classification in each subset, as well as average age and BMI are shown.

## 2.2 Part 1: Choosing tract measurements

For the analysis of the tracts, the ICBM-DTI-81 white-matter labels atlas (**Table S3**) was used, which is composed of 50 tracts. For each of the subject tracts, 12 measurements corresponding to descriptive statistics were calculated (**Table S4**) considering the FA values of the voxels that formed them. Thus, a total of 600 tract measurements were obtained per subject. Subsequently, for each tract measurement, the nonparametric test Wilcoxon rank sum was applied to the tract measurements from the $D_{norm}$ and $D_{over}$ subsets to determine whether there were significant differences ($p_w$ < 0.05) between both. If there was a significant difference, then the respective tract measurement met condition one. Also, for each tract measurement, the Spearman correlation coefficient ($\rho$) and its significance ($p_c$) were calculated between the tract measurements and the BMI of the subjects that formed $D_{all}$. If there was a significant correlation ($p_c$ < 0.05, FDR-corrected), then the respective tract measurement met condition two. Tract measurements that met both conditions were chosen and the total number was called *N*. Boxplots from these tract measurements for the *D*<sub>norm</sub> and *D*<sub>over</sub> sets are shown in the **Figs. S1 to S12**. In the second part of the work, these *N* tract measurements were used to build the two regression models mentioned above.

## 2.3 Part 2: Regression models and creation of biomarkers

Before creating any model, the *D*<sub>all</sub> set was separated into three subsets called *D*<sub>tr</sub>, *D*<sub>val</sub>, and *D*<sub>te</sub>, corresponding to training, validation and testing subsets, and formed by 80, 40, and 40 subjects respectively. Each subset included NW and OB subjects chosen randomly, but under the condition that the three subsets had approximately a homogeneous distribution of the different BMI values available. When possible, the training subset contained twice the number of subjects with the same BMI contained in the validation or testing subsets. The distribution of subjects by BMI in each subset is shown in **Fig. S13**.

### 2.3.1 Model 1: Quantification of BMI

Model 1 aimed to individually predict the obesity level or status of each subject by predicting a quantity for their BMI. The process for creating models by randomly adding and excluding tract measurements was intended to create progressively better models. In **Fig. 1** a diagram schematically representing the process is shown, where the blue shaded area represents the process of adding and the yellow shaded area represents the process of excluding measurements. A step was defined as each occasion on which a new model obtained the best results up to that moment. The proposed process will be described below. From the *N* tract measurements, one was randomly chosen, and a simple linear regression model was created and adjusted considering the subset *D*<sub>tr</sub>. The model was applied to the subjects of *D*<sub>tr</sub> and *D*<sub>val</sub> to predict their BMI. Pearson correlation coefficients $r_{train}$ and $r_{val}$ were calculated by comparing the actual and predicted BMI values. From $r_{train}$ and $r_{val}$, an average coefficient $r_{mean}$ was calculated. This model being the first one created, it was considered the best one up to that point with the coefficient of the highest value, thus having the first step. The change $r_{mean} > r_{mean}^*$, was made, where the asterisk indicated that this coefficient corresponded to the highest so far. From the remaining *N* – 1 measurements, another one was randomly chosen, and together with the first one, a new two-variable linear regression model was created using *D*<sub>tr</sub>. Similarly to before, the new model was applied to the subjects of *D*<sub>tr</sub> and *D*<sub>val</sub>, and the respective coefficients $r_{train}$, $r_{val}$ and $r_{mean}$ were calculated. The following condition was evaluated: $r_{mean} > r_{mean}^*$, where $r_{mean}$ was the coefficient of the last model created and $r_{mean}^*$ was the coefficient of the best model so far. If the condition was met, then the second step was obtained and the following update was made: $r_{mean}^* = r_{mean}$. If the condition was not met, the $r_{mean}^*$ coefficient was not updated, there was no new step and the last measurement added was not considered for the following models to be built.

Subsequently, another measurement was randomly chosen, a new model was created with the previous measurements and the new one chosen was applied to the *D*<sub>tr</sub> and *D*<sub>val</sub> subjects, the respective average coefficient $r_{mean}$ was calculated and the condition $r_{mean} > r_{mean}^*$ was evaluated. This process of randomly searching for a new measurement and, together with the previous ones, creating a new model, was repeated until all the available measurements were considered without repeating any of them. Then, a process similar to the one described above was repeated by searching and adding from two to five measurements at a time. If adding a number of measurements at a time yielded better results, the search was repeated by searching and adding only one measurement at a time.

If the search for five measurements at a time to be added to the best model was exhausted, a different process was carried out in which measurements were excluded from those already added. This consisted of the following. Assuming that the last best model consisted of *m* measurements, then a measurement was randomly chosen from the *m* already added. This measurement was excluded from the model, a new model with *m* – 1 measurements was created, the respective coefficients were calculated and the condition was evaluated in the same way as before. If a new model with better results was created, a new step was obtained and the entire process of searching and adding a single measurement at a time was repeated. If no better results were obtained, the process of excluding a measurement from those already added continued until better results were obtained. If all measurements were exhausted, then a similar process was repeated excluding from two to five measurements at a time. Whenever better results were obtained, the search was restarted and a single measurement was added from those available that had not yet been considered in the best model. Finally, the entire process was concluded when, after adding and excluding up to five measurements at a time and exhausting all available measurements, a new model that yielded better results was not obtained.

Furthermore, due to the random nature of the measurement search, the entire process described was repeated 1,000 times, keeping only the model that obtained the best results from among the 1,000 created, that is, the one that obtained the highest average coefficient $r^*_{mean}$. This was ultimately called model 1. This model was applied to the $D_{tr}$, $D_{val}$ y $D_{te}$ subsets, and from the BMI predictions, the Pearson correlation coefficients $r_{train}$, $r_{val}$, $r^*_{mean}$ y $r_{test}$ were calculated, as well as their $p_p$ values, when compared with the actual BMI values. Correlation graphs and Bland-Altman graphs were created, and the tract measurements and their respective coefficients of the best BMI quantification model were explicitly reported.

### 2.3.2 Model 2: Classification of weight/BMI category

Model 2 aimed to individually predicting the obesity status of subjects by classifying their weight category considering two possibilities: NW and OB. Numerical values equal to -1 and 1 were arbitrarily assigned respectively. The measurements used in the models were also the *N* chosen in the first part. To create model 2, practically the same process of adding and excluding measurements used for model 1 was followed, with the only differences of: Instead of correlation coefficients, the precision of the classifications was calculated. Considering the subsets $D_{tr}$, $D_{val}$ and $D_{te}$, the precisions $a_{train}$, $a_{val}$ and $a_{test}$ were defined respectively. Each one was calculated as the fraction of subjects correctly classified with respect to the total of them in the corresponding subset. To assign a category to the outputs of the models, a threshold equal to 0 was considered such that the outputs lower than the threshold were classified as NW, and higher than the threshold were classified as OB. An average accuracy $a^*_{mean}$ was calculated from the accuracies $a_{train}$ and $a_{val}$, and the condition $a_{mean} > a^*_{mean}$ was tested. The model with the highest $a^*_{mean}$ value was finally called model 2. It was applied to the subsets $D_{tr}$, $D_{val}$ y $D_{te}$, and the accuracies $a_{train}$, $a_{val}$, $a^*_{mean}$ and $a_{test}$ were calculated. Graphs were created to visualize the distribution of classifications. Tract measurements and their respective regression coefficients involved in model 2 were explicitly reported.

### 2.3.3 Creation of biomarkers

Two new obesity biomarkers were proposed. Numerically, each biomarker was constructed from the regression coefficients associated with the tract measurements used in models 1 and 2. Then, each biomarker was reported as a numerical matrix containing the respective coefficients mentioned.

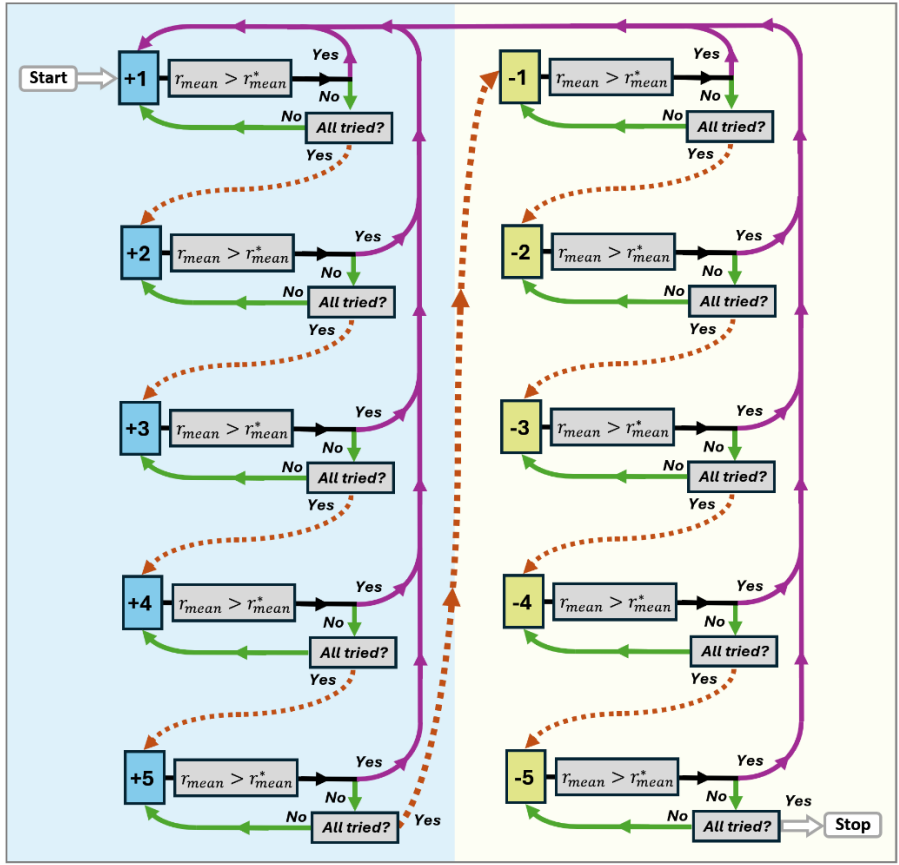

**Figure 1. Process for creating model 1 to quantify BMI.** This process consisted of progressively adding (+) or excluding (-) from 1 to 5 tract measurements at a time to create new and better models. An average correlation coefficient $r_{mean}$ was calculated and the condition $r_{mean} > r_{mean}^*$ was evaluated, with $r_{mean}^*$ being the highest coefficient at the time of evaluation. As long as the condition was met, the aggregation or exclusion was maintained, and the process was restarted (purple solid arrow, Start). Otherwise, aggregation or exclusion was not maintained and the search continued until the available tract measurements were exhausted (green solid arrow), considering from one to five measurements at a time (orange dotted arrow). The entire process ended when no better results were obtained (Stop).

## 3. Results

The tract measurements chosen in the first part are shown in **Fig. 2**. They had significant differences ($p_w < 0.05$) between $D_{norm}$ y $D_{over}$ (condition one) and had significant Spearman correlations ($\rho$) ($p_c < 0.05$, FDR-corrected) with the subjects' BMI when considering $D_{all}$ (condition two). For the tract measurements considered, the $p_w$ values are presented in **Fig. 2**(a), the $p_c$ values in **Fig. 2**(b) and the $\rho$ correlation coefficients in **Fig. 3**(c). Positive and negative correlations can be observed in different measurements in **Fig. 2**(c). Boxplots of the chosen tract measurements for $D_{norm}$ and $D_{over}$ are shown in **Figs. S1** to **S12**. Of the 50 tracts studied, only 14 of them met both conditions mentioned using one or more of the 12 measurements. Also, all 12 measurements were considered in at least one tract. Of the 14 tracts, 9 belonged to the brainstem (middle cerebellar peduncle, pontine crossing tract, left and right corticospinal tract, left and right medial lemniscus, left and right inferior cerebellar peduncle, left superior cerebellar peduncle), 2 were projection tracts (left retrolenticular part of internal capsule, left superior corona radiata), and 3 were association tracts (left external capsule, right cingulate gyrus, right inferior fronto-occipital fasciculus), with 8 of them being homologous tracts and 2 central tracts. In the end, the total number of tract measurements studied was *N* = 71 out of a total of 600 possibilities (= 50 tracts × 12 FA measurements).

For comparison, **Table S5** lists other papers which also reported correlations between FA values of the 14 tracts considered in the present work and BMI, or differences between NW and OB subjects respectively. **Tables S6** and **S7** shows more detailed information on the findings of the listed papers, indicating whether the correlations between FA values and BMI were positive or negative.

From the second part of the work, **Fig. 3**(a) shows a graph with the progress of the steps and the number of measurements used for the creation of the models through the process of adding or excluding measurements. A total of 39 steps were completed (indicating the creation of 39 different models, each obtaining better results than the previous one) by adding or excluding measurements. The last model created obtained the highest $r^*_{mean}$ value using 41 measurements. **Fig. 3**(b) shows the Pearson correlation coefficients $r_{train}$, $r_{val}$, y $r_{test}$ obtained by applying each of the 39 models created to $D_{tr}$, $D_{val}$ and $D_{te}$, in addition to the average coefficient $r^*_{mean}$. **Table 4** shows the results of model 1 for predicting BMI values using the 41 tract measurements.

| Subset | Pearson correlation coefficient | 95% confidence interval | p$_p$ | RMSE | MAPE |
|---|---|---|---|---|---|
| Training ($D_{tr}$) | 0.8124 | (0.7215, 0.8758) | $5.89 \times 10^{-20}$ | 3.22 kg/m$^2$ | 9.92% |
| Validation ($D_{val}$) | 0.8130 | (0.6716, 0.8973) | $1.85 \times 10^{-10}$ | 4.18 kg/m$^2$ | 13.92% |
| Testing ($D_{te}$) | 0.7753 | (0.6116, 0.8754) | $4.23 \times 10^{-9}$ | 3.99 kg/m$^2$ | 12.6% |

**Table 4. Results of model 1 for predicting BMI.** The results obtained after applying the model to predict BMI to the $D_{tr}$, $D_{val}$ and $D_{te}$ subsets are shown, indicating the Pearson correlation coefficient, the 95% confidence interval and the p$_p$ value of the correlation indicating its significance. Their respective root mean square error (RMSE) and mean absolute percentage error (MAPE), for the subjects of $D_{tr}$, $D_{val}$ and $D_{te}$ are presented in the last columns of this Table.

The regression coefficients of model 1 corresponding to the 41 tract measurements are explicitly shown in **Fig. 3**(c). It can be observed that these measurements included the 14 tracts chosen in the first part of the work. **Fig. 3**(d) shows two matrices, one corresponding to the tract measurements of a subject, and the other to the regression coefficients of model 1 associated with the measurements. By performing the element-by-element product summation between the two matrices, the BMI predicted by model 1 is obtained. Numerically, the matrix with the 41 regression coefficients of model 1 associated with measurements of 14 tracts corresponds to the first obesity biomarker proposed in the present work. The correlation graphs and Bland-Altman plots are shown in **Fig. 4**. These results were obtained after applying model 1 to $D_{tr}$, $D_{val}$ and $D_{te}$, obtaining coefficients of reproducibility (RCP) equal to 6.36, 7.56 and 7.87 kg/m$^2$, and coefficients of variation equal to 13%, 16% and 16% respectively.

Similarly to **Fig. 3**, the results obtained during the creation of model 2 are shown in **Fig. 5**. **Fig. 5**(a) shows a graph of the process of adding and excluding measurements. Twenty-two steps in total can be observed, indicating the creation of 22 models. The last of these was model 2, which used 28 tract measurements. **Fig. 5**(b) shows the improvement in the average precision $a^*_{mean}$ during the steps. Also in **Fig. 5**(c) the regression coefficients of model 2 associated with the 28 tract measurements used are explicitly shown. It can be observed that these measurements included 12 of the 14 tracts chosen in the first part (ceasing to consider left superior corona radiata, right inferior fronto-occipital fasciculus). **Fig. 5**(d) shows a diagram of two matrices, one containing the 28 tract measurements of a subject, and the other the correlation coefficients of model 2. The operation between them allowed to classify the weight category of a subject. Numerically, the matrix with the 28 regression coefficients of model 2 associated with measurements of 12 tracts corresponds to the second obesity biomarker proposal in the present work. **Fig. 6** shows graphs with the distribution of the predictions of model 2. **Table 4** shows the results of sensitivity, specificity and accuracy for $D_{tr}$, $D_{val}$ and $D_{te}$, in addition to their root mean square error (RMSE) and mean absolute percentage error (MAPE) values.

The algorithms developed in this work, in addition to the data and results generated, can be freely downloaded for reproducibility in [20].

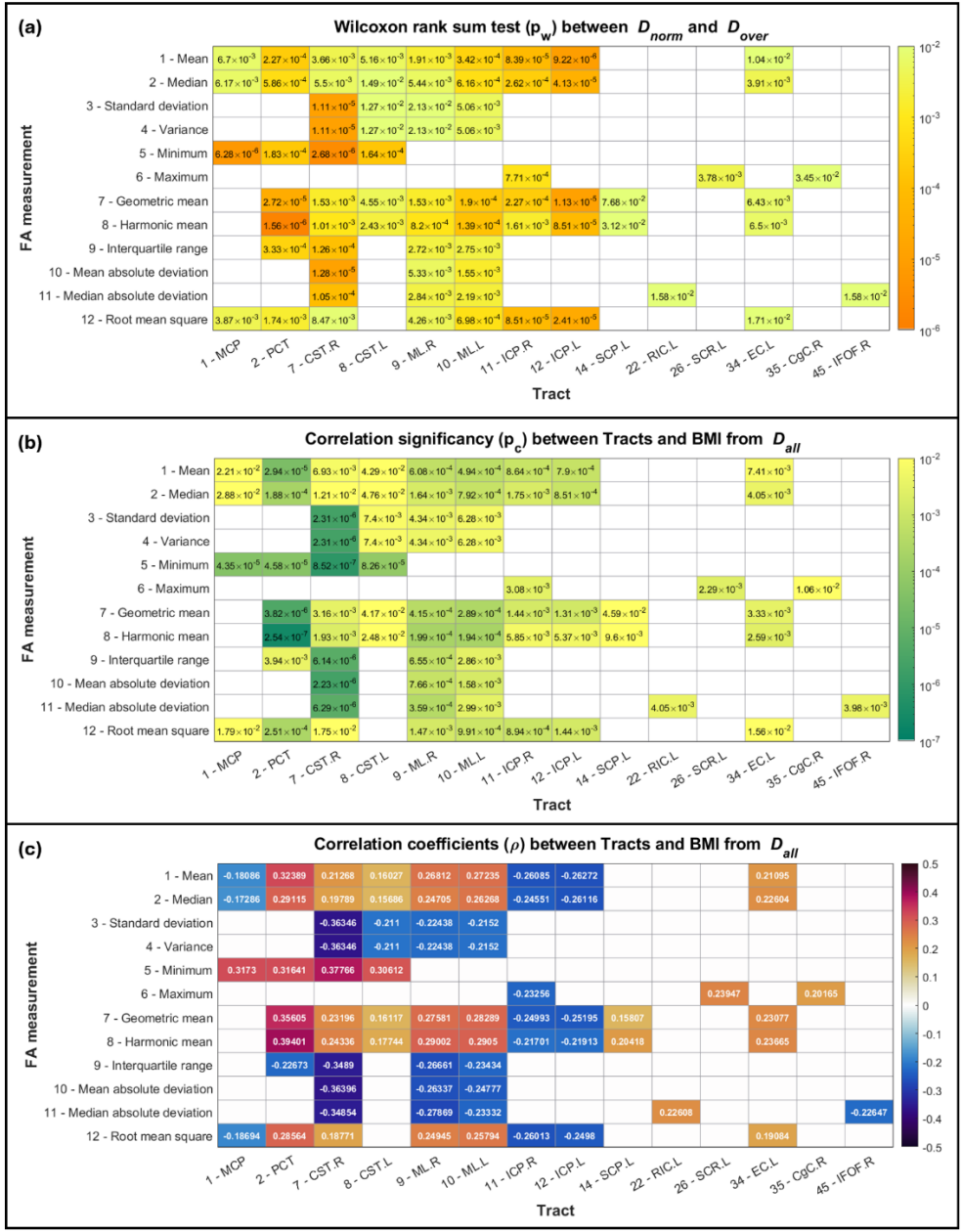

**Figure 2. Tracts and their measurements that met the two conditions of the first part of the study ($p_w < 0.05$ and $p_c < 0.05$, FDR-corrected).** (a) Comparing the data between $D_{norm}$ y $D_{over}$ by applying the Wilcoxon rank sum test, the $p_w$ values of 14 tracts and their respective measurements that met the two conditions are shown. In total, 71 tract measurements were chosen. (b) $p_c$ values indicating the significance ($p_c < 0.05$, FDR-corrected) of the Spearman correlations between the 71 tract measurements and the BMI of the subjects considering $D_{all}$. (c) Spearman correlation coefficients ($\rho$) of the previous tract measurements.

**Figure 3. Results of the process of creating model 1 to quantify BMI.** (a) Number of measurements used and (b) correlation coefficients $r_{train}$, $r_{val}$, $r_{test}$ y $r_{mean}^*$ from each step. (c) Correlation coefficients obtained for model 1, associated with 41 tract measurements (blue cells). (d) Two matrices are shown, one containing the tract measurements of some subject, and another containing the regression coefficients of model 1 shown in (c). Performing the element-wise product summation of the two matrices and adding the model constant $c_q$ = 142.87 kg/m², the BMI predicted can be obtained.

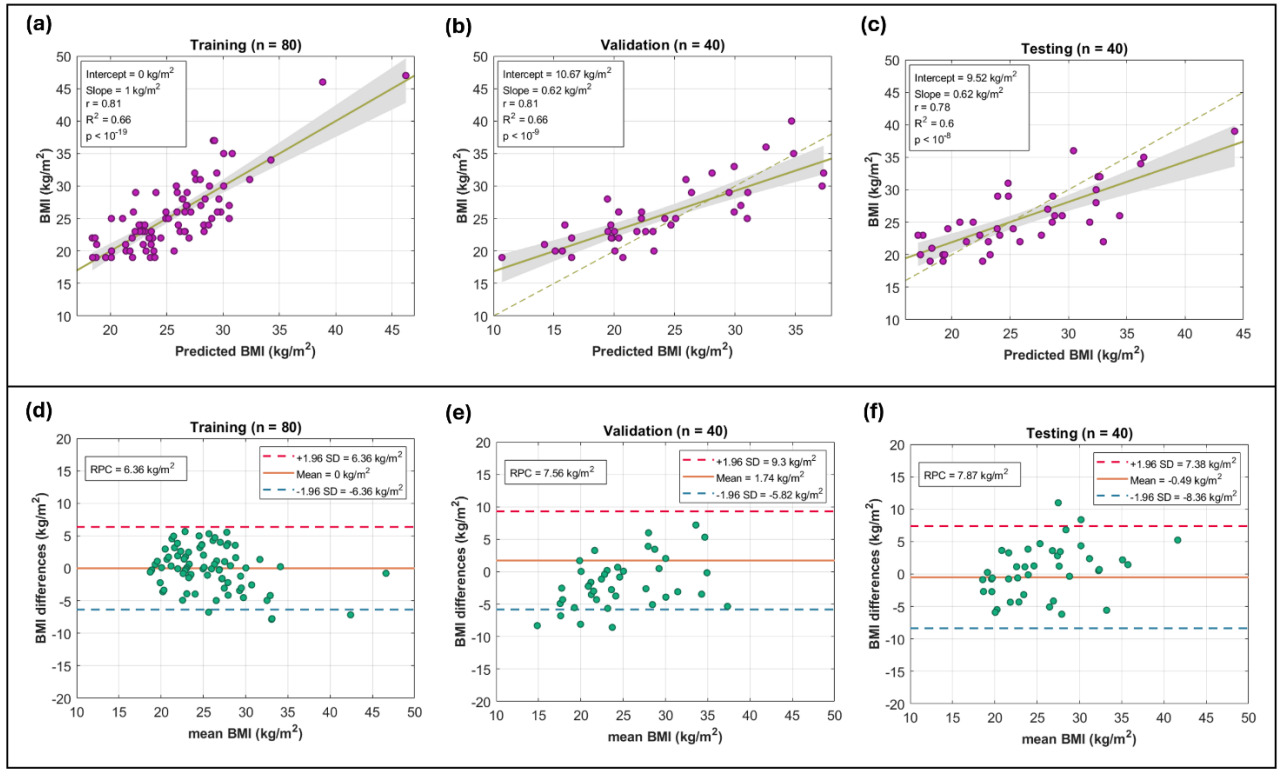

**Figure 4. Correlation and Bland-Altman plots.** Correlation plots are shown in parts (a), (b), and (c), indicating for each subject in the training, validation, and testing subsets ($D_{tr}$, $D_{val}$ and $D_{te}$, respectively), the actual BMI values and those predicted by model 1. Bland-Altman plots are shown in parts (b), (d), and (f). Reproducibility coefficients (RPC), mean and $\pm 1.96$ times the standard deviation (SD) are indicated.

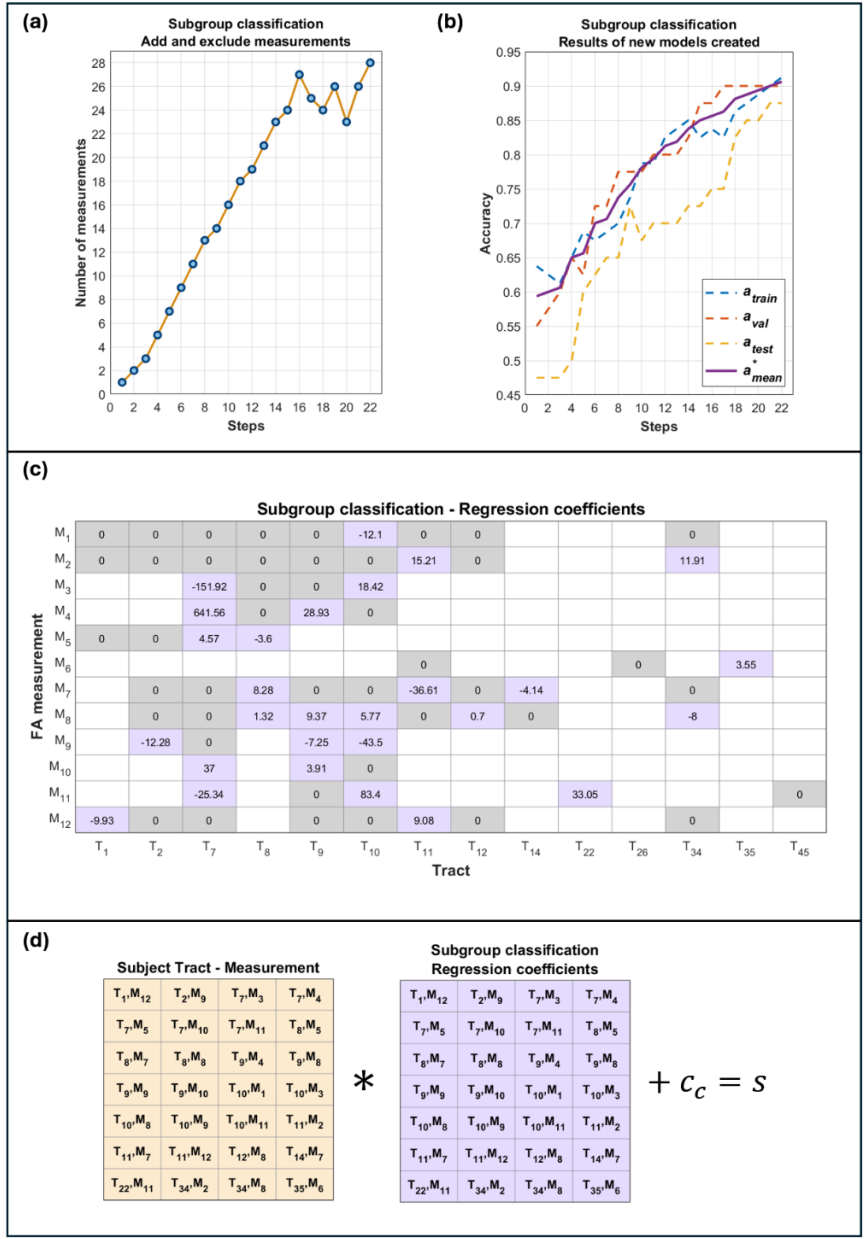

**Figure 5. Results of the process of creating model 2 to classify the weight category of the subjects.** (a) Number of measurements and (b) accuracies $a_{train}$, $a_{val}$, $a_{test}$ and $a^*_{mean}$ from each step. (c) Regression coefficients obtained for the model 2, associated with 28 tract (purple cells). (d) Two matrices are shown, one containing the tract measurements of some subject, and another containing the regression coefficients of model 2 shown in (c). Performing the element-wise product summation of the two matrices and adding the model constant $c_c$ = 5.69, the output $s$ can be obtained to classify the subject considered

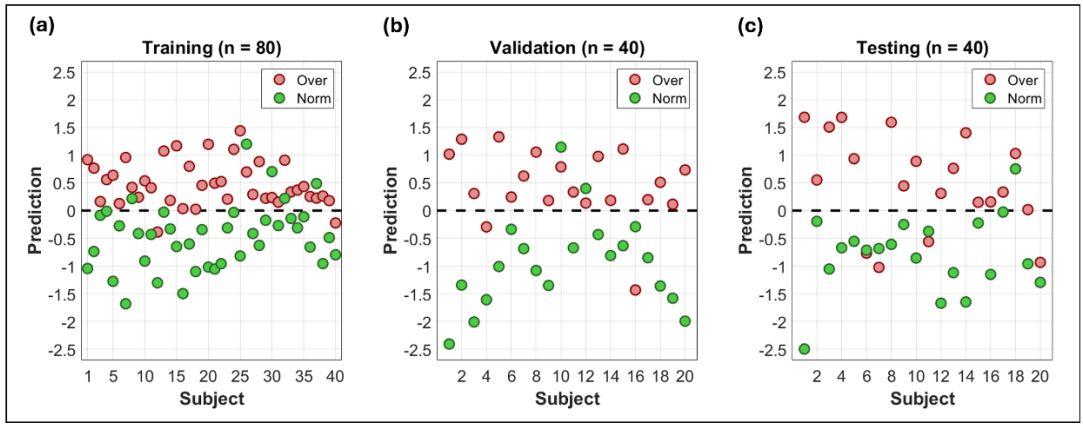

**Figure 6. Plots of classification results**. The distributions of the *s* outputs from model 2 applied to (a) $D_{tr}$, (b) $D_{val}$ and (c) $D_{te}$ respectively are shown. Green circles correspond to subjects with actual normal weight, and red circles to those with actual overweight/obese. A dotted black line at y = 0 indicated the threshold against which subjects were classified. For a subject, if his output was *s* < 0, then that was classified as NW; and if it was *s* > 0, then that was classified as OB.

| Subgroup | Sensibility | RMSE | MAPE | Specificity | RMSE | MAPE | Accuracy | RMSE | MAPE |
|---|---|---|---|---|---|---|---|---|---|
| Training ($D_{tr}$) | 0.875 | 0.707 | 25% | 0.95 | 0.447 | 10% | 0.9125 | 0.591 | 17.5% |
| Validation ($D_{val}$) | 0.9 | 0.633 | 20% | 0.9 | 0.633 | 20% | 0.9 | 0.633 | 20% |
| Testing ($D_{te}$) | 0.95 | 0.447 | 10% | 0.8 | 0.895 | 40% | 0.875 | 0.707 | 25% |

**Table 5. Results of model 2 for subject classification.** Sensitivity, specificity, and accuracy values are shown, along with their respective root mean square error (RMSE) and mean absolute percentage error (MAPE), for the subjects of $D_{tr}$, $D_{val}$ and $D_{te}$.

# 4. Discussion

After studying two sets of subjects (normal weight and overweight/obese), considering specific demographic characteristics, results opposed those usually reported in literature. These consisted of positive correlations between WM integrity and obesity mainly in tracts involved in motor functions, suggesting neuroplasticity in these tracts. Besides, a process for the selection of tract measurements was proposed, such that precise models were built to quantify and classify the obesity level or status of subjects individually. From these, novel brain biomarkers of obesity were proposed, formed by the regression coefficients involved in the models.

## 4.1. WM tract measurements

In this work, the WM tracts were studied through the calculation of descriptive statistics on FA images. All statistics used are well known, although some of them are not usually used to characterize information coming from the WM tracts. The most common statistics used here were the arithmetic mean, median, standard deviation and range [21–25]. In addition to the usual ones, the variance, the harmonic mean, the geometric mean, the interquartile range, maximum, minimum, media absolute deviation, median absolute deviation, and the root of the square mean were used, thus giving a total of 12 statistics. Particularly, in the case of the harmonic mean and the geometric mean, it is known that they are useful for data sets with specific characteristics [26]. The harmonic mean is used instead of the arithmetic mean, when data are expressed as ratios or ranges on different scales. In the case of the geometric mean, it is convenient to use it when there are multiplicative relationships between the data (for example, compound interests). Thus, these expressions of the mean value are used in special situations so that the value obtained is closer to the desired "average." However, there is no explicit prohibition for the use of these statistics when studying data that do not necessarily have the aforementioned characteristics. Furthermore, the geometric and harmonic means were not used looking for a "more correct" average value, but rather they were only used to characterize information from the tracts in a different way. This was also the same reason for using the other statistics already mentioned. As could be seen in **Fig. 3**(b), the central tendency statistics, that is, the (arithmetic) mean, median, geometric mean and harmonic mean, had in common that for the same tract, they all had a significant correlation (positive or negative). On the other hand, based on their definition, there was no linear relationship between them, so although they may represent similar information, they were independent and complementary during the creation of the models 1 and 2. The information extracted calculating the statistics on the FA images proved to be useful to achieve the objectives of the work, since they were part of the selected tract measurements in the models 1 and 2.

## 4.2. Positive correlations between WM integrity and BMI

Many works have reported reduced WM integrity in OB subjects compared to their NW peers. This has been reported as a negative correlation between FA measurements from different tracts and obesity-related measures, such as BMI, waist circumference, fat percentage, etc [5,7,8]. Considering only tract measurements using central tendency statistics (to be able to make comparisons with those obtained by other works), among the 14 reported tracts, 10 of them showed significant correlations with BMI considering the arithmetic mean, median, geometric mean and/or harmonic mean (**Fig. 2**(c)). Among these, 9 major white matter tracts belonged to the brainstem and were involved in motor functions (**Table S4**). From these, 6 showed positive correlations between FA measurements and BMI (pontine crossing tract, left and right corticospinal tracts, left and right medial lemniscus, and left superior cerebellar peduncle). Corticospinal tracts are the major neuronal pathways providing voluntary motor function. They are formed by descending fibers connecting the motor area with the spinal cord to enable distal limb movement [27]. Activity-dependent neuroplasticity of the corticospinal system has been reported, for example during rehabilitation after stroke, where segments in the ipsilesional corticospinal tract showed higher FA values one year after the start of rehabilitation [28]. In a recent work, plasticity of corticospinal tracts was also demonstrated during the early stages of Parkinson's disease (PD) [29]. In that work, longitudinal diffusion tensor imaging before and after the development of prodromal motor PD showed higher FA in corticospinal tract compared to controls, indicating adaptive structural changes in motor networks in concert with nigrostriatal dopamine, as a compensatory process due to the loss of motor skill control. The positive correlations obtained in this work differ from those usually reported by others, where negative correlations were obtained in the mentioned tracts. In view of these findings and following an idea of plasticity, it was hypothesized that the positive correlations found between FA measurements and BMI in the bilateral corticospinal tract and medial lemniscus, as well as the pontine crossing tract, were the result of a plastic process's, probably caused by an increase in the effort to perform voluntary movements of the limbs as a consequence of the difficulty of handling a greater weight. The differences with the results reported by others may originate from the ages of the subjects studied. In this work the age range was restricted from 19 to 26 years, thus being all young adults, to minimize the effect of aging on any brain-related measurements. In other works, however, subjects with a wider age range have been studied, including from young adulthood to adults at older ages. Thus, our results associated obesity in youth with a plasticity effect, possibly lost at older ages. A recently published work was consistent with the plasticity hypothesis presented here. Lv et al. [30] analyzed the effect of cumulative BMI over 16 years on neuroimaging features of brain health in adults of different ages (1,074 adults, 25 to 83 years, 473 females). TBSS analyses were conducted to test between-group differences. WM integrity at the voxel level demonstrated a positive association between cumulative BMI and FA in the pontine crossing tract, middle cerebellar peduncle, and projection fibers (bilateral corticospinal tract). Differences among groups were only significant in young adults (age < 45 years). Based on their findings, the authors suggest that high BMI may be associated with adaptive neuroplasticity or selective neurodegeneration, resulting in higher FA values of projection fibers.

On the other hand, the above hypothesis may be contrary to that reported by another work, which showed that obesity was associated with reduced plasticity in the motor cortex [31]. In that work, 14 subjects (5 females) with obesity and 16 subjects (9 females) with normal weight, aged between 18 and 60 years were studied. A non-invasive brain stimulation protocol known as continuous theta burst transcranial magnetic stimulation (cTBS) was applied to the left motor cortex of the subjects to induce a brief suppression of cortical excitability. The magnitude of suppression of the motor evoked potential was used as a plasticity measurement of the motor cortex. Comparing the response to cTBS between groups demonstrated that there was an impaired plasticity response for the obese group, suggesting that the capacity for plasticity was reduced in people who were obese. It is worth mentioning that that work studied plasticity in the cerebral cortex and not in the WM tracts, in addition to having considered a wider age range and having a reduced number of study subjects. To date and to the best of our knowledge, only three studies have reported a positive correlation between any measure of obesity and WM integrity [10–12]. Thus, these results complement and expand the existing evidence on the impairments in WM integrity in obesity.

Therefore, it can be concluded that the first hypothesis of the study was fulfilled, since comparing NW and OB subjects with specific demographic variables was performed, resulting in opposite findings to those usually reported, and highlighting that the tracts reported in this study corresponded to motor functions.

### 4.3. Quantification and classification models

Based on information from WM tracts, few studies have proposed models to predict the obesity status of subjects individually, either by quantifying the BMI or classifying their weight category [2,3,13,14]. Obviously, it is not useful to quantify or classify the BMI of a subject through complicated and sophisticated methodologies, when this is easily done with conventional instruments. The objective of the models goes in the opposite direction, that is, if a set of tract measurements can predict the obesity status of a subject, then it is hypothesized that these measurements are closely linked to what characterizes the obesity status of the subject. This is intended to provide new knowledge about the effects that obesity causes on WM. In the present work, models 1 and 2 were created to perform the tasks of quantifying the BMI and predicting the weight category respectively. The tract measurements initially considered for their construction were those previously chosen in the first part of the work as they met two conditions: one, having presented significant differences between the subjects that formed $D_{norm}$ and $D_{over}$; and two, to present significant correlations with the subjects' BMI considering $D_{all}$. The tract measurements that ultimately built models 1 and 2 subsequently allowed the proposal of two new biomarkers associated with obesity.

Among the works that have attempted to quantify or classify BMI based on brain characteristics, is the one reported by Byeon et al. [3], who proposed an improvement of the method known as functional correlation tensor, incorporating T1-weighted spatial information. Using the FA images, through the LASSO framework, they identified 26 ROIs from major fiber bundles to predict BMI and to perform classification into three weight subgroups. Predicted BMI showed a mean correlation equal to 0.57, with mean RMSE equal to 4.96±0.65. For classification they obtained an average precision equal to 57.31%. In other work, Park et al. [13] explored a multi-modal approach to predict BMI through connectivity analysis. Significant regions and associated imaging features were identified based on group-wise differences. Using a partial least-square regression (PLSR) framework, their model obtained a correlation coefficient equal to 0.4414, an RMSE equal to 5.26 and SD equal to 5.26. Okudzhava et al. [2] used connectome-based predictive modeling to predict BMI. Multiple linear regression models were built and a leave-one-out cross-validation was used. Their model obtained a correlation coefficient equal to 0.46. Vakli et al. [14] used a convolutional neural network (CNN) for BMI prediction based on T1-weighted structural MRI of the whole brain. The CNN has 230,961 trainable parameters. They used transfer learning to investigate the generalizability of their approach, adapting the model to a different dataset (Images (IXI) dataset). They applied Gradient-weighted Class Activation Mapping (Grad-CAM) to localize brain regions that made a significant contribution to BMI prediction. For the IXI dataset MAE = 3.00 kg/m$^2$; STDAE = 2.12 kg/m$^2$; RMSE = 3.67 kg/m$^2$; Pearson r = 0.49; R$^2$ = 0.21 were obtained.

Therefore, simple models from linear regressions to more complex ones using neural networks have been used to predict obesity status of individual subjects. Simpler models facilitate interpretation, allowing the generation of hypotheses about why measurements in the models are associated with what they predict or classify. More complex models make this task more difficult, since there are a larger number of variables involved that are not trivial to interpret or associate. To reduce the complexity of the models due to the number of variables involved, different processes for model reduction and simplification have been used. In the present work, a simple and original process was proposed to perform this task.

### 4.4. Process of adding and excluding measurements

In the second part of the work, a process was proposed to progressively adding or excluding from 1 to 5 tract measurements at a time to create new and better models. From this process, models 1 and 2 were built to quantify and classify the level or state of obesity of subjects individually, respectively. Other works that proposed models for BMI quantification and/or classification of subjects according to their weight category, used other complex methods for the selection of variables. For example, Byeon et al. [3] used a LASSO framework, which was a regularized regression analysis to select a sparse set of variables that could explain a dependent variable. Park et al. [13], used a partial least-square regression (PLSR) framework, which was a combination of principal component analysis (PCA) and multiple linear regressions. Gupta et al. [32] used sparse partial least squares for discrimination analysis (sPLS-DA). sPLS simultaneously performed variable selection and classification using LASSO penalization. In addition, variable importance in projection (VIP) scores were calculated and a stability analysis was used. It is worth mentioning that there are other proposed models that did not require a variable selection process. Examples of this are the works of Vakli et al. [14] and Finkelstein et al. [33]. In both cases, the models were created from the training, validation and testing of Convolutional Neural Networks (CNNs). CNNs were created specifically for image analysis and had the advantage of automatically learning the necessary image features, without the need of user input. On the other hand, a CNN is characterized by involving a large number of trainable parameters (for example, a total of 230,961 in the work of Vakli et al.), having the characteristic that they cannot have a direct interpretation as in regression models that involve a set of well-identified variables. Since their parameters were not interpretable, other methods have been used to identify the brain regions that contributed most to CNNs for prediction and/or classification tasks. Among them is the Grad-CAM method used by Vakli et al. or the explainable AI (XAI) maps used by Finkelstein et al., which produced localization maps that highlighted regions in the input image (in this case, obtained from the brain) that were important for prediction and/or classification.

Unlike other works, the proposed process for measurement selection can be partially considered as a "brute force algorithm", since it was a simple and straightforward approach in which almost all possible combinations were tested randomly. However, for model 1, adding the condition $r_{mean} > r_{mean}^*$, reduced the search space and improved efficiency. The methodologies for variable selection in other works are mainly useful when there is a large number of features, so overfitting is desired to be avoided by reducing the complexity of the model. In our case, the number of possible features was relatively low. Considering initially 50 tracts and 14 statistical measurements, 600 tract measurements were available. Then, in the first part, applying the two required conditions, this number was reduced to 71. From this number of measurements, the simple and novel process described for choosing variables was applied, and two models were created, one to quantify BMI using 42 tract measurements (from 11 statistics measured in some of 13 tracts), and another for classification using 28 tract measurements (from 12 statistics measured in some of 12 tracts). Therefore, the proposed models were characterized by using a relatively small number of measurements before and after the measurement selection process. In addition, it required a small amount of computational resources during its application, since it started from one measurement onwards, while the other methodologies mentioned started from the total number of measurements and advanced by reducing the number of them by applying more complex and demanding processes from the beginning. Although other reported methods were more complex and statistically robust, in the present work the results obtained using the proposed process allowed obtaining high correlations with the real BMI values by applying model 1, and high precision in the classification of weight categories by applying model 2, thus demonstrating its effective and convenient utility for the problem addressed.

### 4.5. Brain biomarkers associated with obesity

Brain biomarkers allow the evaluation of individual information extracted from multimodal analysis methodologies. This is useful as clinical tools to help with the diagnosis, monitoring and prognosis of diseases, response to drugs, or otherwise, to establish objective and quantitative associations between differential characteristics between study groups [34,35]. Therefore, these biomarkers have potential clinical applications, in addition to being useful for the generation of new knowledge that allows directing novel research based on their findings. Among the most studied brain biomarkers are those that describe the difference between chronological age and predicted age based on structural and functional neuroimaging data. These have been associated with neurodegenerative diseases, such as schizophrenia, Alzheimer's, Parkinson's, multiple sclerosis, or with an increase in patient mortality [36–38]. Regarding brain biomarkers associated with obesity, there are those studying quantitative MRI metrics relating obesity with neuroinflammation in brain regions involved in food intake [39], with negative emotional states [40], or with dysfunction in executive regions causing cognitive deficit [41]. In the present work, two novel biomarkers that associated obesity with decreased or increased integrity of several WM tracts were proposed. Regarding the tracts that presented increased integrity, it is known that these are related to motor functions. The proposed biomarkers were represented as fixed value matrices, formed by the regression coefficients of models 1 and 2, and based on well-known but rarely used statistical measurements to characterize WM integrity. These biomarkers were interesting findings, since part of them were constructed from information on an increased integrity of motor tracts, also suggesting plasticity and adaptation processes that differ from what is usually reported by other works. The possible main reason for such differences lies in the specific demographic characteristics of the study group formed by young right-handed women with a medium level of education. This was different from what is usually studied by other works, in which they analyze wide ranges of ages and genders mixed with the objective of generalizing their results.

Considering all the above, it can be concluded that the second hypothesis of the work was fulfilled. Still, the final objective of the present work was that the findings had an impact on clinical applications for the identification of risk factors associated with brain alterations, or new mechanisms of action for the prevention and treatment of these changes [7]. These biomarkers were an initial suggestion that sought to direct future research based on the reported findings. After validation by independent works studying larger samples, the usefulness of these biomarkers associated with obesity can be validated.

### 4.6 Limitations

Limitations of the present work include the small sample size and the lack of generalizability of the results due to the restriction of demographic characteristics. On the other hand, the type of analysis performed based on VBA has known methodological limitations, particularly the insufficient specificity of the measured parameters and the limited accuracy of WM tract reconstruction, because DTI has little capacity to differentiate crossing fibers within a voxel [9]. This may consequently imply that FA alterations may reflect different underlying factors such as axonal loss or injury, changes in myelin content, inflammation, or shifts in extracellular and intracellular water concentrations [2]. Interpreting findings in a biological context remains a challenge. Finally, there is always a limitation in correlation studies on the causal interpretation of the results, since it cannot be concluded whether the brain alterations are caused by obesity, or whether obesity causes those brain alterations.

In this work, BMI was the only measurement associated with obesity. Considering that this index is more sensitive to the lean mass of the subjects than to their body and abdominal fat content, other measurements associated with adiposity should be used. These could include waist measurement, percentage of adipose tissue, amount of subcutaneous fat, amount of visceral fat, among others. The proposed biomarkers should also be applied to other databases formed by subjects with the same anthropometric characteristics considered in this study to validate the findings and prove their reproducibility. Other study groups, formed by older age ranges including men, should be analyzed to extend and compare results dependent on age and gender. In this work, the use of an original and simple process for the selection of variables for regression models was proposed, so that other statistically more robust methods to identify multicollinearities between variables could be implemented (such as the well-known LASSO and Ridge regressions). Also stability-based quantification metrics could be considered based on simple cross-validations. Finally, in addition to multiple linear regression models, other models can be considered based on CNNs, combined with novel interpretation methods to identify features that have a greater contribution to the prediction models.

## 5. Conclusions

Novel brain biomarkers associated with obesity were sought by studying tract measurements based on well-known but rarely used statistics to characterize FA images. In the present work, subjects with specific demographic variables were studied, such as being right-handed, women aged between 19 and 26 years, and with a medium level of education. This was done to avoid variations in brain measurements dependent on other confounds such as age and gender. After choosing tract measurements that showed significant differences between the two groups formed by subjects with NW and OB respectively, and whose measurements were also correlated with the subjects' BMI, a simple and novel process was applied to choose the measurements that would allow the creation of models to quantify and classify the obesity status of individual subjects. The biomarkers sought were then constructed from the regression coefficients of the models. Among the tract measurements involved in the models, several were found to show positive correlations between WM integrity and obesity, mainly in tracts related to motor functions. This was contrary to what is usually reported. The findings allowed us to hypothesize a possible neuroplasticity in motor tracts related to obesity. As future work, the process carried out should be repeated in larger databases, including men, as well as a wider age range, to obtain results that can be generalized. Finally, the proposed novel obesity biomarkers allow the generation of new knowledge, and have the ultimate goal of subsequently being a useful tool in the clinical environment for the prevention and treatment of WM changes due to obesity.

## 6. Contribution statement

**José Gerardo Suárez-García**: Conceptualization, Methodology, Software, Validation, Formal analysis, Investigation, Writing – Original Draft – Review & Editing. **María Isabel Antonio-de la Rosa**: Software, Investigation, Review & Editing. **Nora Coral Soriano-Becerril**: Software, Investigation, Review & Editing. **Javier M. Hernández López:** Software, Investigation, Review & Editing. **Benito de Celis-Alonso**: Conceptualization, Formal analysis, Resources, Data Curation, Writing - Review & Editing, Supervision, Project administration, Funding acquisition.

## Declaration of Competing Interest

The authors declare that they have no known competing financial interests or personal relationships that could have appeared to influence the work reported in this paper.

## Data Availability Statement

Data from the 160 subjects studied in this work are available in the ID1000 database contained in the open-access database Amsterdam Open MRI Collection (AOMIC) available at [15]. The algorithms developed and the data generated can be freely downloaded at [20].

## Acknowledgment

The author José Gerardo Suárez-García was supported by the National Council of Sciences, Technologies and Humanities (CONAHCYT) to carry out this work, through a posdoctoral scholarship.

# Supplementary figures and tables

| sub-0001 | sub-0089 | sub-0170 | sub-0308 | sub-0422 | sub-0489 | sub-0636 | sub-0712 |
| --- | --- | --- | --- | --- | --- | --- | --- |
| sub-0002 | sub-0091 | sub-0192 | sub-0309 | sub-0425 | sub-0515 | sub-0640 | sub-0727 |
| sub-0006 | sub-0094 | sub-0204 | sub-0354 | sub-0429 | sub-0536 | sub-0644 | sub-0804 |
| sub-0037 | sub-0095 | sub-0216 | sub-0360 | sub-0431 | sub-0539 | sub-0664 | sub-0811 |
| sub-0047 | sub-0104 | sub-0226 | sub-0363 | sub-0443 | sub-0547 | sub-0671 | sub-0833 |
| sub-0072 | sub-0118 | sub-0254 | sub-0368 | sub-0446 | sub-0566 | sub-0687 | sub-0834 |
| sub-0079 | sub-0127 | sub-0264 | sub-0374 | sub-0448 | sub-0578 | sub-0691 | sub-0849 |
| sub-0084 | sub-0141 | sub-0279 | sub-0384 | sub-0458 | sub-0579 | sub-0700 | sub-0875 |
| sub-0086 | sub-0145 | sub-0286 | sub-0385 | sub-0467 | sub-0617 | sub-0709 | sub-0882 |
| sub-0088 | sub-0169 | sub-0291 | sub-0412 | sub-0477 | sub-0628 | sub-0711 | sub-0917 |

**Table S1. Subjects with normal weight studied in this work.** Participant IDs provided in the ID1000 database.

| sub-0008 | sub-0133 | sub-0221 | sub-0298 | sub-0451 | sub-0572 | sub-0764 | sub-0850 |
| --- | --- | --- | --- | --- | --- | --- | --- |
| sub-0020 | sub-0138 | sub-0236 | sub-0301 | sub-0483 | sub-0582 | sub-0774 | sub-0857 |
| sub-0026 | sub-0143 | sub-0243 | sub-0302 | sub-0487 | sub-0591 | sub-0775 | sub-0859 |
| sub-0040 | sub-0154 | sub-0259 | sub-0312 | sub-0494 | sub-0643 | sub-0794 | sub-0862 |
| sub-0064 | sub-0155 | sub-0260 | sub-0342 | sub-0495 | sub-0646 | sub-0809 | sub-0866 |
| sub-0099 | sub-0162 | sub-0278 | sub-0357 | sub-0499 | sub-0649 | sub-0818 | sub-0870 |
| sub-0121 | sub-0165 | sub-0283 | sub-0381 | sub-0508 | sub-0666 | sub-0826 | sub-0886 |
| sub-0123 | sub-0177 | sub-0285 | sub-0404 | sub-0525 | sub-0688 | sub-0827 | sub-0893 |
| sub-0130 | sub-0187 | sub-0292 | sub-0405 | sub-0535 | sub-0726 | sub-0845 | sub-0902 |
| sub-0131 | sub-0190 | sub-0297 | sub-0437 | sub-0563 | sub-0753 | sub-0847 | sub-0927 |

**Table S2. Subjects with overweight/obesity studied in this work.** Participant IDs provided in the ID1000 database.

**Figures S1 - S12. Boxplots of the tract measurements studied.** Boxplots for the $D_{norm}$ (blue) and $D_{over}$ (red) sets, with the white matter tract indicated on the horizontal axis and the value of the measured statistic on the vertical axis. Only the tracts and measurements that had statistically significant differences ($p_w < 0.05$) after applying the nonparametric Wilcoxon rank sum test, comparing the subjects of the $D_{norm}$ and $D_{over}$ sets, are shown. It is worth mentioning that these tracts and measurements are the same ones that obtained significant Spearman correlations ($p_c < 0.05$, FDR-corrected) with the BMI values of all the study subjects of the $D_{all}$ set, which can be observed in Fig. 2.

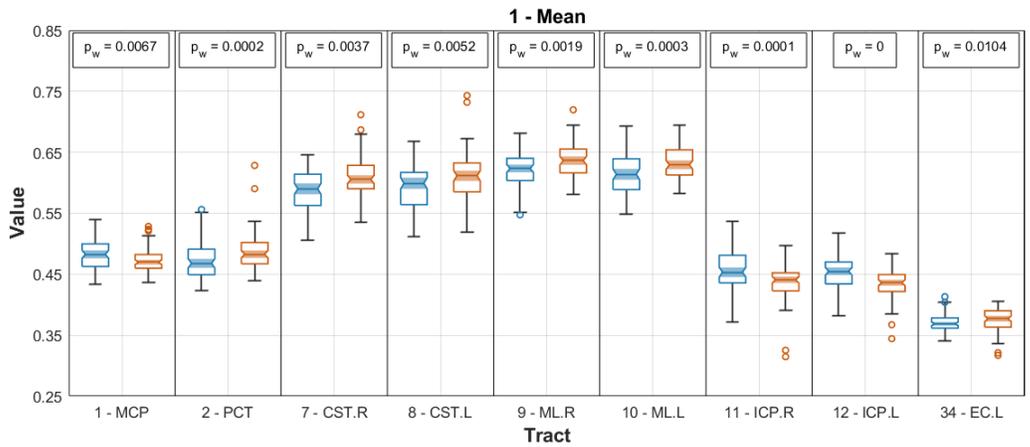

**Figure S1.**

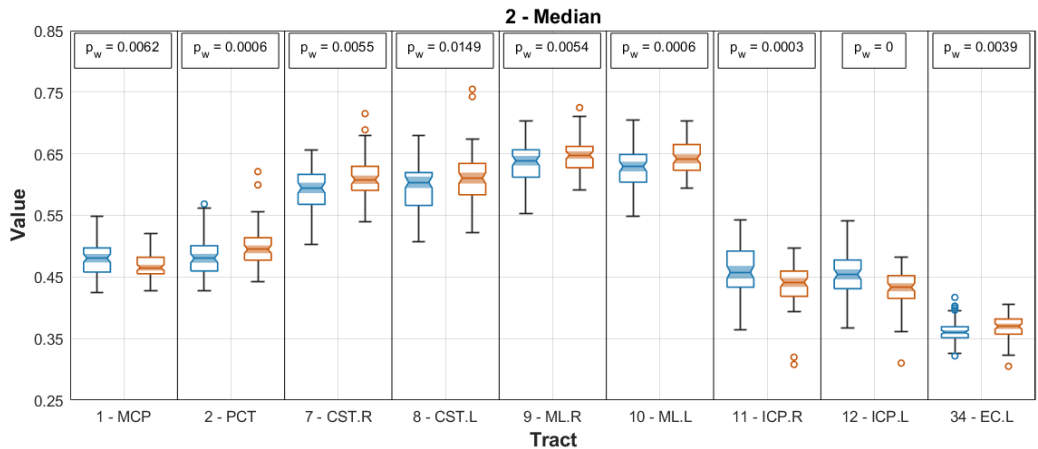

**Figure S2.**

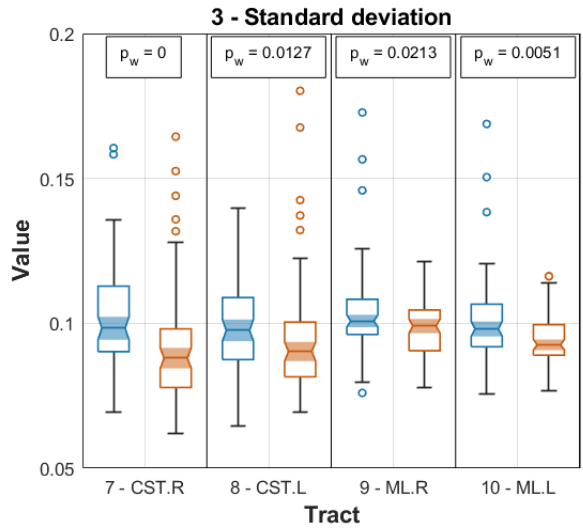
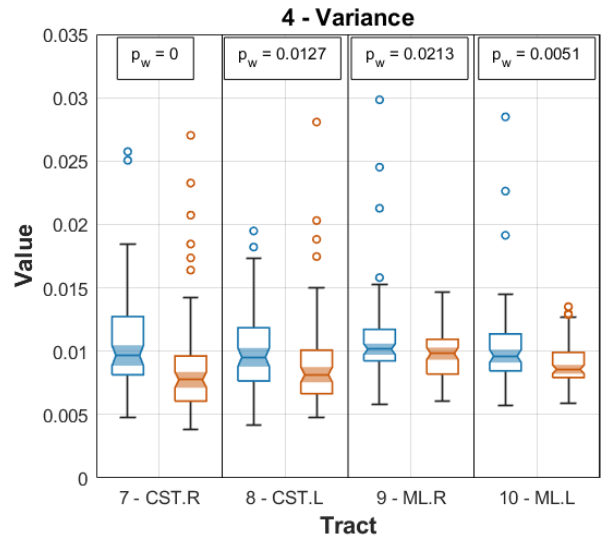

**Figures S3 and S4.**

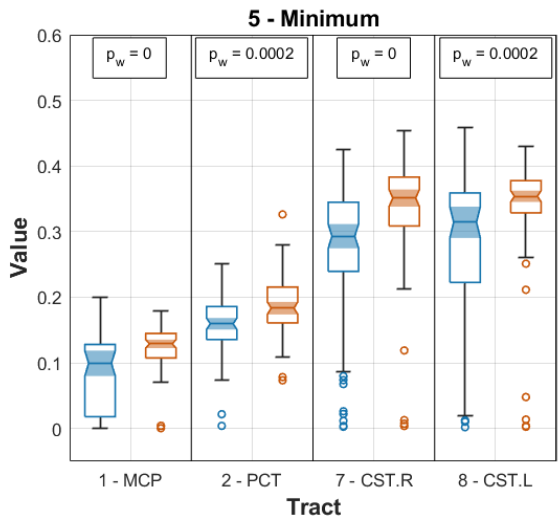
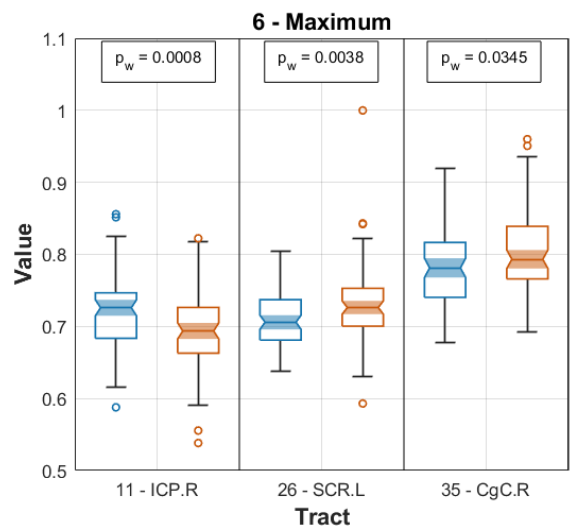

**Figures S5 and S6.**

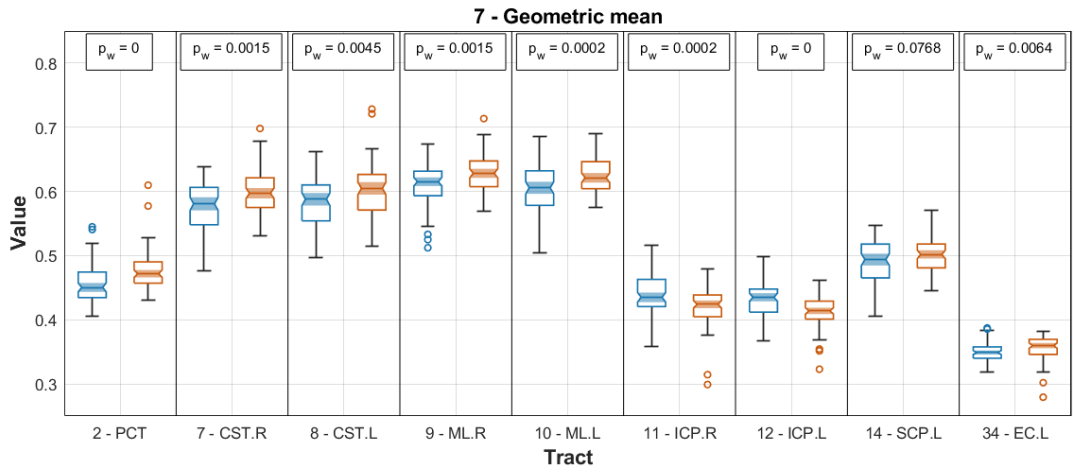

**Figure S7.**

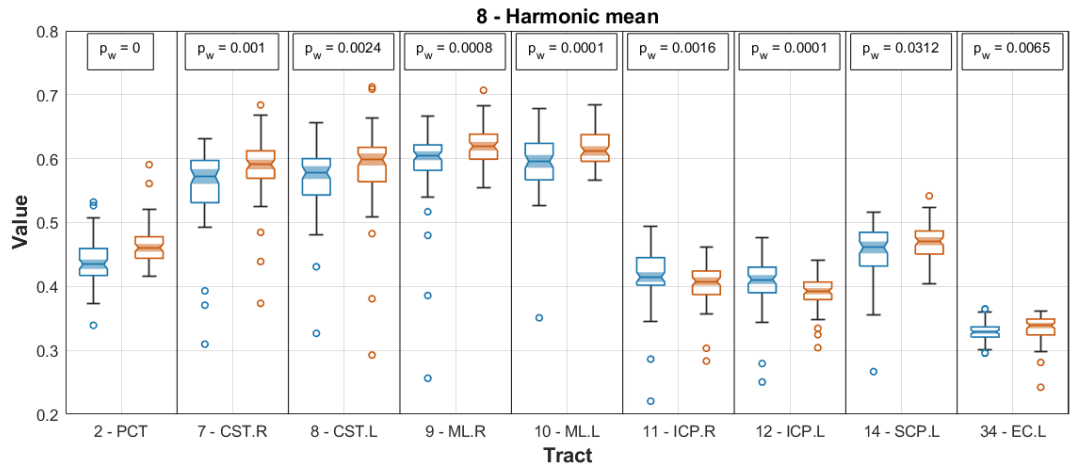

**Figure S8.**

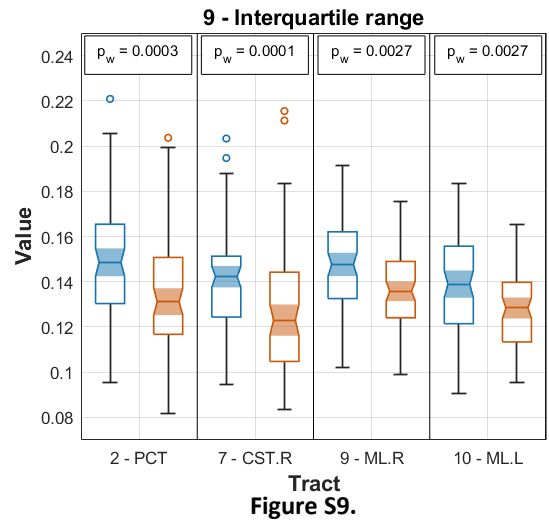

**Figure S9.**

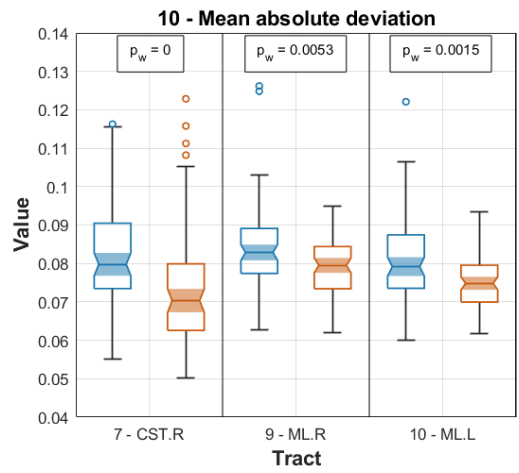

**Figure S10**

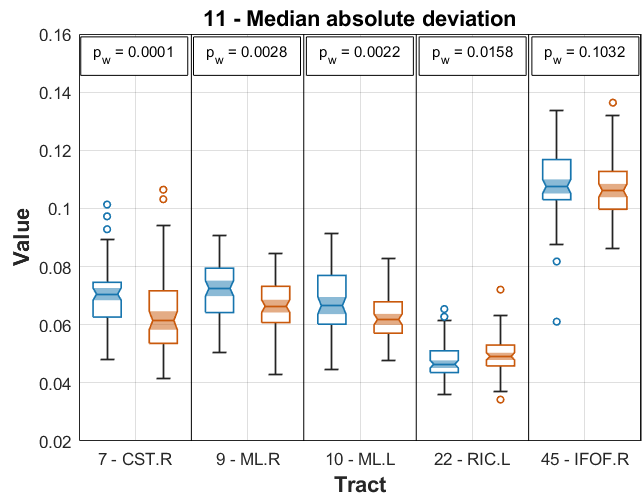

**Figure S11.**

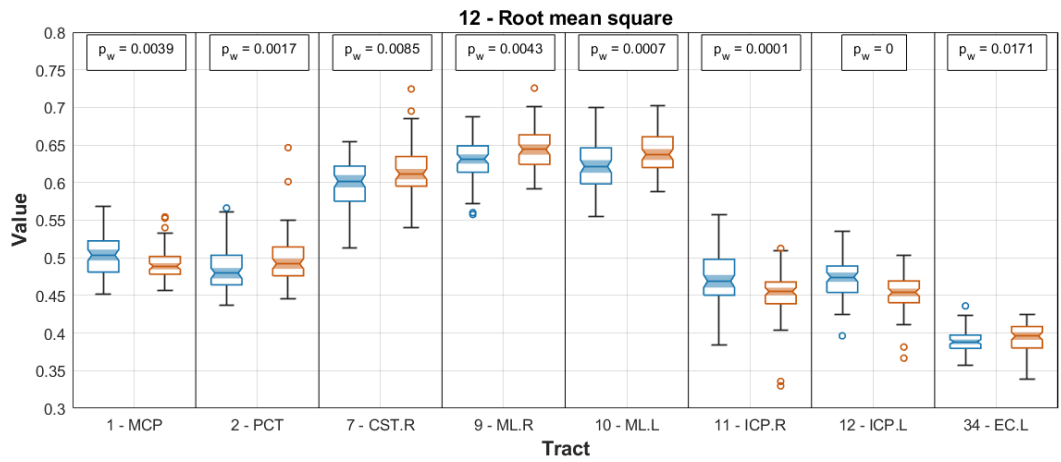

**Figure S12.**

| Index | Tract | Abbreviation | Category |
|---|---|---|---|
| 1 | Middle cerebellar peduncle | MCP | Brainstem |
| 2 | Pontine crossing tract | PCT | Brainstem |
| 3 | Genu of corpus callosum | gCC | Commissural |
| 4 | Body of corpus callosum | bCC | Commissural |
| 5 | Splenium of corpus callosum | sCC | Commissural |
| 6 | Fornix (column and body of fornix) | FX | Association |
| 7 | Corticospinal tract R | CST.R | Brainstem |
| 8 | Corticospinal tract L | CST.L | Brainstem |
| 9 | Medial lemniscus R | ML.R | Brainstem |
| 10 | Medial lemniscus L | ML.L | Brainstem |
| 11 | Inferior cerebellar peduncle R | ICP.R | Brainstem |
| 12 | Inferior cerebellar peduncle L | ICP.L | Brainstem |
| 13 | Superior cerebellar peduncle R | SCP.R | Brainstem |
| 14 | Superior cerebellar peduncle L | SCP.L | Brainstem |
| 15 | Cerebral peduncle R | CP.R | Projection |
| 16 | Cerebral peduncle L | CP.L | Projection |
| 17 | Anterior limb of internal capsule R | ALIC.R | Projection |
| 18 | Anterior limb of internal capsule L | ALIC.L | Projection |
| 19 | Posterior limb of internal capsule R | PLIC.R | Projection |
| 20 | Posterior limb of internal capsule L | PLIC.L | Projection |
| 21 | Retrolenticular part of internal capsule R | RIC.R | Projection |
| 22 | Retrolenticular part of internal capsule L | RIC.L | Projection |
| 23 | Anterior corona radiata R | ACR.R | Projection |
| 24 | Anterior corona radiata L | ACR.L | Projection |
| 25 | Superior corona radiata R | SCR.R | Projection |
| 26 | Superior corona radiata L | SCR.L | Projection |
| 27 | Posterior corona radiata R | PCR.R | Projection |
| 28 | Posterior corona radiata L | PCR.L | Projection |
| 29 | Posterior thalamic radiation R | PTR.R | Projection |
| 30 | Posterior thalamic radiation L | PTR.L | Projection |
| 31 | Sagittal stratum R | SS.R | Association |
| 32 | Sagittal stratum L | SS.L | Association |
| 33 | External capsule R | EC.R | Association |
| 34 | External capsule L | EC.L | Association |
| 35 | Cingulum (cingulate gyrus) R | CgC.R | Association |
| 36 | Cingulum (cingulate gyrus) L | CgC.L | Association |
| 37 | Cingulum (hippocampus) R | CgH.R | Association |
| 38 | Cingulum (hippocampus) L | CgH.L | Association |
| 39 | Fornix (cres) / Stria terminalis R | FX/ST.R | Association |
| 40 | Fornix (cres) / Stria terminalis L | FX/ST.L | Association |
| 41 | Superior longitudinal fasciculus R | SLF.R | Association |
| 42 | Superior longitudinal fasciculus L | SLF.L | Association |
| 43 | Superior fronto-occipital fasciculus R | SFOF.R | Association |
| 44 | Superior fronto-occipital fasciculus L | SFOF.L | Association |
| 45 | Inferior fronto-occipital fasciculus R | IFOF.R | Association |
| 46 | Inferior fronto-occipital fasciculus L | IFOF.L | Association |
| 47 | Uncinate fasciculus R | UF.R | Association |
| 48 | Uncinate fasciculus L | UF.L | Association |
| 49 | Tapetum R | TAP.R | Commissural |
| 50 | Tapetum L | TAP.L | Commissural |

**Table S3. ICBM-DTI-81 white-matter labels atlas.** White matter parcellation.

| Measurement | Equation | Measurement | Equation |
|---|---|---|---|
| $M_1$: Mean | $\mu(A) = \frac{1}{N}\sum_{i=1}^{N} A_i$ | $M_7$: Geometric mean | $GM(A) = \left[\prod_{i=1}^{N} A_i\right]^{\frac{1}{N}}$ |
| $M_2$: Median | $\tilde{\mu}(A)$ | $M_8$: Harmonic mean | $HM(A) = N \Big/ \sum_{i=1}^{N} \frac{1}{A_i}$ |
| $M_3$: Standard desviation | $SD(A) = \sqrt{\frac{1}{N-1}\sum_{i=1}^{N}|A_i - \mu(A)|^2}$ | $M_9$: Interquartile range | $IQ(A) =$ Third quartile$(A)$ $-$ first quartile$(A)$ |
| $M_4$: Variance | $VAR(A) = \frac{1}{N-1}\sum_{i=1}^{N}|A_i - \mu(A)|^2$ | $M_{10}$: Mean absolute deviation | $MAD(A) = \mu(|A_i - \mu(A)|)$ |
| $M_5$: Minimum | $min(A)$ | $M_{11}$: Median absolute deviation | $MDAD(A) = \tilde{\mu}(|A_i - \tilde{\mu}(A)|)$ |
| $M_6$: Maximum | $max(A)$ | $M_{12}$: Root mean square value | $RMS(A) = \sqrt{\frac{1}{N}\sum_{i=1}^{N}|A_i|^2}$ |

**Table S4. Tract measurements.** For each tract, 12 descriptive statistics were calculated from the voxel values considering the FA images.

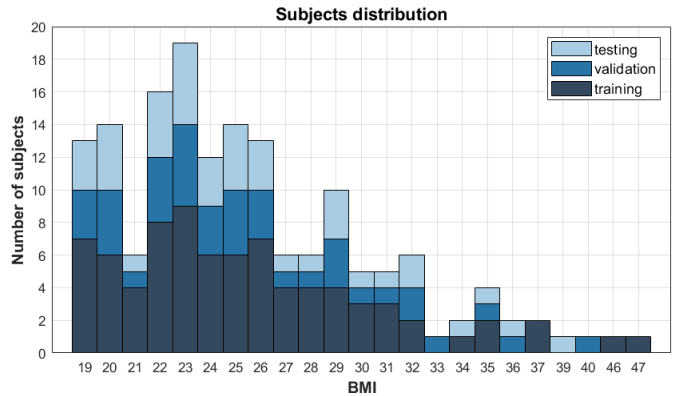

**Figure S13. Distribution of subjects by subset and BMI.** For the training, validation and testing subsets, 80, 40 and 40 subjects respectively were randomly selected. An approximate 2:1 ratio was attempted between subjects with the same BMI value in the training subset and the validation or testing subsets.

| Study | Subjects (Female) | Age (years) | BMI (kg/m$^2$) | DTI analysis |
|---|---|---|---|---|
| Best et al. 2020 [42] | 1065 (575) | 28.75 ± 3.67 | 26.40 ± 5.11 | TBSS |
| Birdsill et al. 2017 [11] | 168 (96) | 49.5 ± 6.4 | 29.8 ± 6.7 | TBSS |
| Dietze et al. 2023 [7] | 5,237 | 8 to 92 | 27.29 ± 2.91 | VBA/TBSS |
| Figley et al. 2016 [43] | 32 (16) | 18 to 57 | 18 to 37 | VBA |
| Karlsson et al. 2013 [44] | 22 (15) (N) 23 (18) (OV/OB) | 46.4±9.5 (N) 47.3±8.9 (OV/OB) | 24.0±2.3 (N)  43.2±3.7 (OV/OB) | VBA |
| Kullmann et al. 2016 [45] | 48 (23) | 21 to 36 | 19.5 to 39.3 | VBA |
| Lou et al. 2014 [46] | 22 (27) | 29.04±7.32 (N), 31.72±8.04 (OV/OB) | 21.54±2.06 (N)  31.44±3.34 (OV/OB) | TBSS |
| Papageorgiou et al. 2016 [47] | 120 (76) (N) 148 (77) (OV/OB) | 39.8 ± 15.8 (N) 51.5 ±14.8/ 52.0 ± 15.7 (OV/OB) | Not reported | TBSS |
| Patel et al. 2022 [48] | 281 (128) | 60.9 ± 9.6 | 26.5 ± 4.5 | TBSS |
| Rahmani et al. 2022 [12] | 231 (124) | 69.2±8.3 | 27.3±4.7 (male)  28.2±6.6 (female) | CT |
| Repple et al. 2018 [4] | 369 (186) | 39.4 (11.2) | 24.7 (4.1) | TBSS |
| Shott et al. 2015 [49] | 24 (24) (N) 18 (18) (OV/OB) | 27.4±6.3 (N),  28.7±8.3 (OV/OB) | 21.64 ± 1.26 (N)  34.78 ± 4.44 (OV/OB) | VBA |
| Verstynen et al. 2012 [10] | 155 (77) | 40.7 ± 6.2 | 27.15 ± 4.82 | VBA |
| Zhang et al. 2018 [50] | 636 (322) | 55.43 ± 16.05 | 25.82 ± 3.69 | TBSS |

**Table S5. Summary of related work.** Details on published papers that have reported correlations between WM integrity and obesity.

| WM Tract | Category | Function/connection | Statistics | FA Correlation/Contrast (This work) | Study | FA Correlation/Contrast |
|---|---|---|---|---|---|---|
| 1 - Middle cerebellar peduncle | Brainstem | Initiation, planning, and timing of volitional motor activity; posture, balance, and coordination [51]. | Mean, median | Negative BMI, N > OV/OB | Dietze et al. | Negative BMI |
| | | | | | Kullmann et al. | Negative BMI |
| | | | | | Verstynen et al. | Positive BMI |
| 2 - Pontine crossing tract | Brainstem | Part of middle cerebellar peduncle [51]. | Mean, median, geometric, armonic | Positive BMI, N < OV/OB | Best et al. | Negative BMI |
| | | | | | Verstynen et al. | Negative BMI |
| 7 - Corticospinal tract R | Brainstem | Principal motor pathway for voluntary movements [27]. | Mean, median, geometric, armonic | Positive BMI, N < OV/OB | Rahmani et al. | Negative BMI |
| | | | | | Karlsson et al. | N > OV/OB |
| | | | | | Dietze et al. | Negative BMI |
| | | | | | Verstynen et al. | Negative BMI |
| | | | | | Papageorgiou et al. | N > OV/OB |
| | | | | | Lou et al. | Negative BMI/WC |
| 8 - Corticospinal tract L | Brainstem | Principal motor pathway for voluntary movements [27]. | Mean, median, geometric, armonic | Positive BMI, N < OV/OB | Rahmani et al. | Negative BMI (male) |
| | | | | | Rahmani et al. | Positive BMI (female) |
| | | | | | Patel et al. | Negative AFR |
| | | | | | Best et al. | Negative BMI |
| | | | | | Karlsson et al. | N > OV/OB |
| | | | | | Dietze et al. | Negative BMI |
| | | | | | Verstynen et al. | Negative BMI |
| | | | | | Papageorgiou et al. | N > OV/OB |
| | | | | | Lou et al. | Negative BMI/WC |
| 9 - Medial lemniscus R | Brainstem | Convey sensations of touch, vibration, proprioception, and 2-point discrimination [51]. | Mean, median, geometric, armonic | Positive BMI, N < OV/OB | Verstynen et al. | Negative BMI |
| 10 - Medial lemniscus L | Brainstem | Convey sensations of touch, vibration, proprioception, and 2-point discrimination [51]. | Mean, median, geometric, armonic | Positive BMI, N < OV/OB | Verstynen et al. | Negative BMI |
| 11 - Inferior cerebellar peduncle R | Brainstem | Motor control such as coordination of movement control of balance, posture, and gait [52]. | Mean, median, geometric, armonic | Negative BMI, N > OV/OB | Verstynen et al. | Negative BMI |
| 12 - Inferior cerebellar peduncle L | Brainstem | Motor control such as coordination of movement control of balance, posture, and gait [52]. | Mean, median, geometric, armonic | Negative BMI, N > OV/OB | Verstynen et al. | Negative BMI |
| 14 - Superior cerebellar peduncle L | Brainstem | Motor coordination and balance network [51]. | geometric, armonic | Positive BMI, N < OV/OB | Verstynen et al. | Negative BMI |
| | | | | | Zhang et al. | Negative BMI |
| 34 - External capsule L | Association | Carries fibers directly to the striatum from the prefrontal cortex [53]. | Mean, median, geometric, armonic | Positive BMI, N < OV/OB | Patel et al. | Negative AFR |
| | | | | | Birdsill et al. | Positive WC |
| | | | | | Shott et al. | N > OV/OB |
| | | | | | Repple et al. | Negative BMI/WC |
| | | | | | Zhang et al. | Negative BMI |

**Table S6. Comparison of findings.** The tracts that presented significant correlations ($p_c < 0.05$, FDR-corrected) in this work, between WM integrity and BMI, are listed. The statistics mean (arithmetic), median, geometric mean and/or harmonic mean calculated on the tracts in the FA images were considered. It is indicated whether the correlations were positive (blue cells) or negative (pink cells), or if there were differences between the study groups. The works that reported findings in the listed tracts are also indicated.

| WM Tract | Category | Function/connection | Statistics | FA Correlation/Contrast (This work) | Study | FA Correlation/Contrast |
|---|---|---|---|---|---|---|
| 22 Retrolenticular part of internal capsule L | Projection | Contains fibers of the optic radiations [54]. | Median absolute deviation | Positive BMI, N < OV/OB | Birdsill et al. | Positive WC |
| 26 Superior corona radiata L | Projection | Projects from the thalamus to the sensory cortex [55]. | Maximum | Positive BMI, N < OV/OB | Figley et al. | Negative BFP |
| | | | | | Patel et al. | Negative AFR |
| | | | | | Birdsill et al. | Positive WC |
| | | | | | Verstynen et al. | Negative BMI |
| | | | | | Shott et al. | N > OV/OB |
| 35 Cingulum (cingulate gyrus) R | Association | Affect, visceromotor control; response selection in skeletomotor control; visuospatial processing and memory access [51]. | Maximum | Positive BMI, N < OV/OB | Birdsill et al. | Positive WC |
| | | | | | Verstynen et al. | Negative BMI |
| | | | | | Papageorgiou et al. | N > OV/OB |
| 45 Inferior fronto-occipital fasciculus R | Association | Integration of auditory and visual association cortices with prefrontal cortex [51]. | Median absolute deviation | Negative BMI, N > OV/OB | Rahmani et al. | Negative BMI (male) |
| | | | | | Rahmani et al. | Positive BMI (female) |
| | | | | | Figley et al. | Negative BFP |
| | | | | | Karlsson et al. | N > OV/OB |
| | | | | | Papageorgiou et al. | N > OV/OB |
| | | | | | Repple et al. | Negative BMI/WC |

**Table S7. Comparison of findings considering other statistics.** Information similar to Table S4 is shown, but considering other statistics used in this work.